\documentclass[aps,prx,reprint,longbibliography]{revtex4-2}
\usepackage{amsmath,amssymb,bm}
\usepackage{graphicx}
\usepackage{physics}
\usepackage{hyperref}
\hypersetup{colorlinks, linkcolor={blue}, citecolor={blue}, urlcolor={blue}}
\usepackage{xcolor}

\begin{document}

\title{Collective Charge-\(2e\) Bosonic Excitations in Charge-Ordered Systems}

\author{Ping Tang}
\email{tang.ping.a2@tohoku.ac.jp}
\affiliation{Institute for Materials Research, Tohoku University, Sendai 980-8577, Japan}
\date{\today}
\begin{abstract}
Charge order is conventionally characterized by a static modulation of the
electronic density, while its collective excitations remain less developed
from a quasiparticle perspective than those of spin-ordered systems. Here, we
formulate the collective excititations of charge order within an effective
charge-pseudospin model, in which the charge-ordered ground state is represented
by staggered pseudospin order. Quantizing fluctuations around this ordered state
via a Holstein--Primakoff transformation, we reveal two branches of bosonic
quasiparticles with degenerate, gapped dispersions that carry opposite
quantized electric charges $\pm 2e$. We therefore term these charge-$2e$
bosons ``\textit{bichargons},'' closely paralleling the two magnon branches
of a bipartite antiferromagnet that carry opposite spin angular momenta.
We show that the diffusion of thermally excited bichargons under a temperature
gradient generates a net charge current when the degeneracy between the two
oppositely charged branches is lifted by tuning the chemical potential away
from the charge-neutrality point, thereby giving rise to a bichargon Seebeck effect. In contrast, despite carrying finite electric charges, a thermal bichargon
gas remains electrically insulating under a static electric field because
bichargon quasiparticle number is not conserved. An ac electric field, however, can
parametrically generate coherent pairs of oppositely charged bichargons that
support a finite dc drift current in the presence of a bias electric field,
offering a bosonic analogue of photoconductivity mediated by photoexcited
electron--hole pairs in semiconductors. Our results establish bichargons as a
new class of collective bosonic charge carriers in charge-ordered systems and
extend the quasiparticle paradigm of magnons in spin-ordered systems to the
charge sector, with electric charge replacing spin angular momentum as the
transported quantity.
\end{abstract}

\maketitle

\section{Introduction}
The emergence of order in condensed matter is often associated with spontaneous symmetry breaking, while its low-energy collective excitations can be represented by quasiparticles that inherit fundamental characteristics of the underlying order and may function like real particles~\cite{Goldstone1962,Anderson1972}. Magnetism emerges from spin order that breaks time-reversal symmetry. The collective excitations of spin order can be quantized into bosonic quasiparticles known as magnons, which carry elementary spin angular momentum and an associated magnetic dipole moment. In contrast to electrons, magnons transport pure spin angular momentum through magnetic insulators without accompanying charge motion~\cite{PhysRevLett.109.096603,Cornelissen2015,PhysRevB.94.014412,PhysRevB.93.060403,li2016observation,lebrun2018tunable}, providing charge-neutral spin carriers for low-dissipation spintronic devices~\cite{kruglyak2010magnonics,Chumak2015}. Ferroelectricity, on the other hand, is characterized by the spontaneous ordering of electric dipoles that breaks spatial inversion
symmetry~\cite{xu2013ferroelectric}. The quasiparticle excitations of ferroelectric order, termed ferrons~\cite{Tang2022,Bauer2022,PhysRevApplied.20.050501,Tang2024,
choe2026observation,zhang2026electric,shen2025observation,itoh2026observation}, constitute the electric counterparts
of magnons by carrying an electric dipole moment, enabling a variety of unconventional thermoelectric
transport~\cite{PhysRevLett.126.187603,TangPRL2022,wooten2023electric,
PhysRevB.107.L121406,3y1m-66s1,itoh2026observation} and dynamical
phenomena~\cite{Tang2024,choe2026observation,zhang2026electric,jana2026ferron}. More recently, multiferrons~\cite{gm7n-8ftp,2sj3-33ky}, which combine the
properties of magnons and ferrons by possessing both spin angular momentum and
an electric dipole moment, have been proposed as quasiparticles associated
with the collective precessional dynamics of ferroelectric order.

Charge order represents one of the most ubiquitous forms of electronic ordering in correlated materials and is characterized by a spatial modulation of the electronic density that breaks translational symmetry. The non-equilibrium dynamics of charge order have been extensively explored in response to external electrical or optical stimuli, including depinning and sliding~\cite{PhysRevB.19.3970,RevModPhys.60.1129,monceau2012electronic}, melting~\cite{fiebig2000sub,PhysRevLett.98.097402,PhysRevLett.103.155702,PhysRevLett.105.187401}, and transitions to other electronic phases~\cite{stojchevska2014ultrafast,geremew2019bias}. Recently, an electric-current-driven reversal of charge-order polarity was
reported in LuFe$_2$O$_4$~\cite{kikkawa2026electronic}, closely paralleling
current-induced magnetization switching in spin-ordered
systems~\cite{slonczewski1996current,PhysRevB.54.9353,ralph2008spin}.
An effective pseudospin model that describes the charge-ordered state in
terms of staggered pseudospin order~\cite{tang2026pseudospin} reveals that
the reversal process follows Landau--Lifshitz--Gilbert-like pseudospin
dynamics analogous to those of a bipartite antiferromagnet, establishing
a close correspondence between charge- and spin-order dynamics. However,
while spin order supports magnons carrying spin angular momentum, it remains
an open question whether staggered charge-pseudospin order can support
analogous quasiparticle excitations carrying electric charge itself.

\begin{figure}
    \centering
    \includegraphics[width=8.6cm]{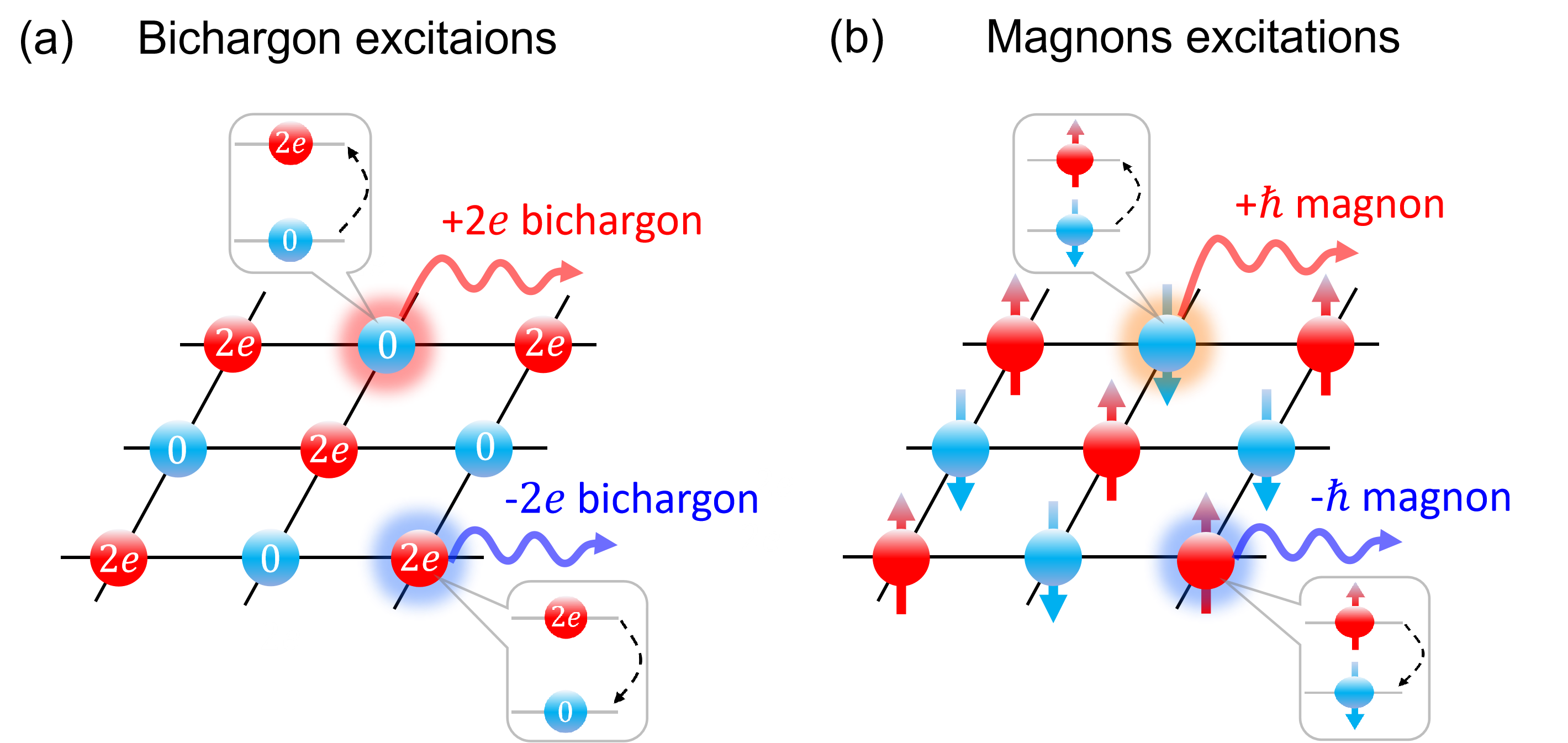}
    \caption{Schematic of (a) two oppositely charged bichargon excitations in a
charge-ordered state with alternating doubly occupied and empty sites,
arising from electron-pair transitions between the two local charge
configurations, and (b) their magnon counterparts in a bipartite antiferromagnet, associated with spin-flip excitations and carrying opposite spin angular momenta $\pm\hbar$.}
    \label{fig-1}
\end{figure}

In this work, we provide an affirmative answer to this question by quantizing
the collective pseudospin excitations of charge order via a Holstein--Primakoff
transformation. We find two degenerate, gapped bosonic branches at the
charge-neutrality point that carry opposite quantized electric charges
$\pm 2e$, which we term \emph{bichargons}, as illustrated in
Fig.~\ref{fig-1}. Tuning the chemical potential away from the charge-neutrality
point generates a uniform pseudospin field that shifts the two bichargon
branches in opposite directions, closely paralleling the magnetic-field-induced
splitting of magnon branches in antiferromagnets.

As bosonic quasiparticles, bichargons exhibit both similarities to and
fundamental distinctions from conventional electronic charge carriers in
thermoelectric transport. We show that a temperature gradient drives the
diffusion of thermally excited bichargons, generating a charge Seebeck
response when the populations of the oppositely charged branches are
imbalanced, analogous to the magnonic spin Seebeck effect in
antiferromagnets~\cite{PhysRevB.87.014423,PhysRevLett.116.097204,
PhysRevB.93.014425}. Despite carrying quantized electric charge, however,
a thermal bichargon gas \textit{cannot} support steady dc electrical
conduction under a static electric field. This absence of electrical
conduction results from the exact compensation of the field-driven drift
current by a diffusion current arising from the field-induced inhomogeneous
local-equilibrium distribution of bichargons, highlighting their unusual
character as charged bosonic quasiparticles whose quasiparticle number is
not conserved.

Beyond thermoelectric transport, we show that, in charge-ordered states that
also break inversion symmetry, a static electric field can tune the bichargon
dispersions via dipolar coupling, whereas an ac electric field above a
threshold can parametrically excite coherent pairs of oppositely charged
bichargons with opposite momenta. Although each generated bichargon pair
carries zero net charge and momentum, its oppositely charged constituents
can support a finite dc charge current under an applied bias electric field,
in sharp contrast to thermally excited bichargons. This mechanism provides
a bosonic counterpart of photoconductivity mediated by photoexcited
electron--hole pairs in semiconductors, with parametrically excited
bichargons serving as the mobile charge carriers.

The remainder of this paper is organized as follows. In Sec.~II, we introduce
the charge-pseudospin model and identify bichargons as the collective bosonic
excitations of charge order. In Sec.~III, we investigate the responses of
bichargons to temperature gradients and static and ac electric fields. In
Sec.~IV, we discuss bichargons in relation to other collective excitations
and nonequilibrium phenomena in correlated electron systems and summarize
our main conclusions.

\section{Pseudospin model and collective excitations}
The charge order in correlated electron systems can be captured by the extended Hubbard model incorporating both on-site ($U$) and intersite ($V$) Coulomb interaction~\cite{PhysRevLett.53.2327,PhysRevB.39.9397,PhysRevLett.106.236805,PhysRevLett.110.166401,PhysRevLett.111.036601,PhysRevB.95.115149,PhysRevB.99.245146},
\begin{align}
H=&-t\sum_{\langle ij\rangle,\sigma}\left(c_{i\sigma}^{\dagger}c_{j\sigma}+\mathrm{H.c.}\right)
-\mu\sum_i n_i \nonumber\\&+U\sum_i n_{i\uparrow}n_{i\downarrow}+V\sum_{\langle ij\rangle}n_i n_j,
\label{eq:Hubbard}
\end{align}
where $t$ denotes the nearest-neighbor hopping amplitude, $\mu$ the chemical potential, $c_{i\sigma}^{(\dagger)}$ the annihilation (creation) operator for an electron with spin $\sigma$ at site $i$, $n_{i\sigma}=c_{i\sigma}^{\dagger}c_{i\sigma}$, and $n_i=n_{i\uparrow}+n_{i\downarrow}$ the total electron-number operator at site $i$. At half filling, the on-site interaction $U$ favors a Mott state
with singly occupied sites and antiferromagnetic correlations, whereas the
intersite interaction $V$ favors charge disproportionation between
neighboring sites. In the atomic limit ($t=0$), the condition $U<ZV$
stabilizes a charge-ordered ground state, where $Z$ is the lattice
coordination number. We focus on this charge-ordered regime, in which the
relevant low-energy local configurations are the empty state $|0\rangle_i$
and the doubly occupied state
$|2\rangle_i=c_{i\uparrow}^{\dagger}c_{i\downarrow}^{\dagger}|0\rangle_i$ that alternate between neighboring sites, while the singly occupied states
enter only as virtual intermediate states. Projecting Eq.~(\ref{eq:Hubbard}) onto the low-energy charge-order subspace yields an effective pseudospin-$1/2$ Hamiltonian~\cite{tang2026pseudospin},
\begin{align}
H_{\rm eff}
=&
\sum_{\langle ij\rangle}
\left[
J_z\tau_i^z\tau_j^z
-
\frac{J_\perp}{2}
\left(
\tau_i^+\tau_j^-+\tau_i^-\tau_j^+
\right)
\right]
-h_z\sum_i\tau_i^z ,
\label{eq:Hps}
\end{align}
where $\tau_i^z$ and $\tau_i^{\pm}=\tau_i^x\pm i\tau_i^y$ denote the
longitudinal and ladder components of the pseudospin operator
$\boldsymbol{\tau}_i=(\tau_i^x,\tau_i^y,\tau_i^z)$, respectively, with
\begin{equation}
\tau_i^z=\frac{n_i-1}{2},\qquad
\tau_i^+=c_{i\uparrow}^{\dagger}c_{i\downarrow}^{\dagger},\qquad
\tau_i^-=c_{i\downarrow}c_{i\uparrow}.
\label{eq:tauz_charge}
\end{equation}
Here, $J_z\simeq4(V+t^2/\Delta)$ favors Ising-type staggered pseudospin order, whereas $J_\perp\simeq4t^2/\Delta$ gives rise to a ferromagnetic-like transverse coupling associated with coherent pair transfer between neighboring sites; $\Delta$ denotes the excitation energy of the virtual singly occupied configurations. The uniform field $h_z=2\mu-U-2ZV$ is conjugate to the total longitudinal
pseudospin and hence to the total electron number. At the
particle-hole-symmetric half-filled point, $\mu=U/2+ZV$ and thus $h_z=0$; a finite $h_z$ therefore parametrizes deviations from this symmetric
charge-neutrality point. Since $J_z>J_\perp>0$ for $V>0$, the ground state
develops staggered pseudospin order along the $z$ axis, corresponding to a
charge-ordered configuration with alternating doubly occupied and empty sites.

\subsection{Collective bichargon excitations}
We now address the collective bichargon excitations by quantizing the pseudospin fluctuations around the
charge-ordered ground state. For the equilibrium staggered pseudospin
configuration $\langle\tau_A^z\rangle=\tau_0$ and
$\langle\tau_B^z\rangle=-\tau_0$, where $A$ and $B$ label the two pseudospin
sublattices, the linearized Holstein--Primakoff transformations read
\begin{subequations}
\begin{align}
\tau_{iA}^{z}&=\tau_0-a_i^\dagger a_i,\,\,
\tau_{iA}^{+}\simeq\sqrt{2\tau_0}\,a_i,\,\, \tau_{iA}^{-}\simeq\sqrt{2\tau_0}\,a_i^{\dagger}\\
\tau_{jB}^{z}&=-\tau_0+b_j^\dagger b_j,\,\,
\tau_{jB}^{+}\simeq\sqrt{2\tau_0}\,b_j^\dagger,\,\,\tau_{iB}^{-}\simeq\sqrt{2\tau_0}\,b_i
\end{align}
\label{eq:HP}
\end{subequations}
where $a_i^{(\dagger)}$ and $b_j^{(\dagger)}$ are bosonic operators
representing pseudospin fluctuations on the $A$ and $B$ sublattices,
respectively. Substituting Eq.~(\ref{eq:HP}) into Eq.~(\ref{eq:Hps}) and
Fourier transforming to momentum space yields the quadratic bosonic
Hamiltonian
\begin{align}
H_{\rm eff}
=&\sum_{\mathbf{k}}
\Big[
(\mathcal{A}+h_z)a_{\mathbf{k}}^\dagger a_{\mathbf{k}}
+
(\mathcal{A}-h_z)b_{\mathbf{k}}^\dagger b_{\mathbf{k}}
\nonumber\\
&+\mathcal{B}_{\mathbf{k}}
\left(
a_{\mathbf{k}}b_{-\mathbf{k}}
+
a_{\mathbf{k}}^\dagger b_{-\mathbf{k}}^\dagger
\right)
\Big]\nonumber\\
=&
\sum_{\mathbf{k}}
\left[(\varepsilon_{\mathbf{k}}+h_{z})
\alpha_{\mathbf{k}}^\dagger\alpha_{\mathbf{k}}
+(\varepsilon_{\mathbf{k}}-h_{z})
\beta_{\mathbf{k}}^\dagger\beta_{\mathbf{k}}
\right] +\text{const.}
\label{eq:Hpm_quad}
\end{align}
where $\mathcal{A}=Z\tau_0J_z$, $\mathcal{B}_{\mathbf{k}}
=-Z\tau_0J_\perp\gamma_{\mathbf{k}}$, and $\gamma_{\mathbf{k}}
=1/Z\sum_{\bm{\delta}_{ij}}
e^{i\mathbf{k}\cdot\bm{\delta}_{ij}}$ is the nearest-neighbor structure factor, with $\bm{\delta}_{ij}$ denoting the
nearest-neighbor bond vectors. In the second equality, we have diagonalized the quadratic Hamiltonian via
the Bogoliubov transformation
$a_{\mathbf{k}}
=u_{\mathbf{k}}\alpha_{\mathbf{k}}
-v_{\mathbf{k}}\beta_{-\mathbf{k}}^\dagger$
and
$b_{-\mathbf{k}}^\dagger
=u_{\mathbf{k}}\beta_{-\mathbf{k}}^\dagger
-v_{\mathbf{k}}\alpha_{\mathbf{k}}$,
where
$u_{\mathbf{k}}
=\sqrt{(\mathcal{A}+\varepsilon_{\mathbf{k}})/(2\varepsilon_{\mathbf{k}})}$
and
$v_{\mathbf{k}}
=\operatorname{sgn}(\mathcal{B}_{\mathbf{k}})
\sqrt{(\mathcal{A}-\varepsilon_{\mathbf{k}})/(2\varepsilon_{\mathbf{k}})}$.
Here,
$\varepsilon_{\mathbf{k}}=\sqrt{\mathcal{A}^2-\mathcal{B}_{\mathbf{k}}^2}$
is the common dispersion of the two degenerate bichargon eigenmodes
$\alpha$ and $\beta$ at $h_z=0$. For a $d$-dimensional hypercubic lattice with lattice
constant $a$, the long-wavelength expansion
$\gamma_{\mathbf{k}}\simeq1-k^2a^2/(2d)$ yields
\begin{equation}
\varepsilon_{\mathbf{k}}
\simeq
\sqrt{\Delta_b^2+\hbar^2c_{b}^2k^2}, \label{disp}
\end{equation}
where $\Delta_b=Z\tau_0\sqrt{J_z^2-J_\perp^2}$ is the bichargon
excitation gap at $h_z=0$, while $c_b=Z\tau_0J_\perp a/(\hbar\sqrt{d})$
is the characteristic velocity parameter.

Eq.~(\ref{disp}) exhibits a gapped relativistic-like dispersion, analogous
to that of magnons in an easy-axis antiferromagnet~\cite{rezende2020fundamentals,
rezende2019introduction}. The longitudinal field $h_z$, which can be induced
by doping away from half filling, acts as a charge-sector counterpart of a
Zeeman field, lifting the degeneracy of the two bichargon branches through
opposite energy shifts $\pm h_z$. Since $h_z$ is conjugate to the total
electron number, the opposite responses of the two branches to $h_z$ suggest
that they carry opposite electric-charge quantum numbers, as we demonstrate
explicitly below. When $|h_z|$ reaches the zero-field bichargon gap
$\Delta_b$, the lower branch softens at $\mathbf{k}=0$, signaling an
instability of the collinear charge-ordered state. Beyond this critical field,
the system develops a uniform transverse pseudospin component, marking the
breakdown of the collinear charge-ordered state as it is tuned sufficiently
far from the particle-hole-symmetric point. In the following, we focus on the
regime $|h_z|\ll\Delta_b$, where the collinear charge-ordered state
remains stable.

\subsection{Quantized electric charge of bichargons}
While magnons in spin-ordered systems carry quantized spin angular momentum,
we show that bichargons carry quantized electric charges of $\pm2e$ as
bosonic quasiparticle excitations of charge order. The fluctuations of the total electric charge relative to half filling due
to bichargon excitations are
\begin{align}
Q
&\equiv
-e\sum_i(n_i-1)
=
-2e\sum_i\tau_i^z
\nonumber\\
&=
2e\sum_{\mathbf{k}}
\left(
\alpha_{\mathbf{k}}^\dagger\alpha_{\mathbf{k}}
-
\beta_{\mathbf{k}}^\dagger\beta_{\mathbf{k}}
\right),
\label{eq:localcharge}
\end{align}
where $e>0$ is the elementary charge. The electric charge carried by each
bichargon follows directly from the commutators of its creation operator with
the total charge operator,
\begin{align}
[Q,\alpha_{\mathbf{k}}^\dagger]
&=
+2e\,\alpha_{\mathbf{k}}^\dagger,
&
[Q,\beta_{\mathbf{k}}^\dagger]
&=
-2e\,\beta_{\mathbf{k}}^\dagger,
\label{eq:chargecommutator}
\end{align}
showing that a single $\alpha$ ($\beta$) bichargon carries the quantized
electric charge $+2e$ ($-2e$). At $h_z=0$, the two oppositely charged bichargon branches are degenerate
with equal thermal populations,
$\langle\alpha_{\mathbf{k}}^\dagger\alpha_{\mathbf{k}}\rangle
=\langle\beta_{\mathbf{k}}^\dagger\beta_{\mathbf{k}}\rangle$,
yielding $\langle Q\rangle=0$. Doping away from half filling generates a
finite $h_z$, which lifts the degeneracy of the two branches and leads to
unequal thermal populations and hence a finite $\langle Q\rangle$. 

The quantization of the bichargon charge follows from the axial $U(1)$ pseudospin symmetry of Eq.~(\ref{eq:Hps}) about the $z$ axis, i.e., $[\sum_i\tau_i^z, H_{\text{eff}}]=0$. Since $\sum_i\tau_i^z$ is directly related to the total electron number, this symmetry expresses conservation of the total electronic charge. In this respect, the two bichargon branches closely resemble the two magnon branches of a uniaxial bipartite antiferromagnet carrying opposite quantized spin angular momenta $\pm\hbar$~\cite{rezende2020fundamentals,rezende2019introduction}, with electric charge replacing spin angular
momentum as the conserved quantum number, as schematically shown in Fig.~\ref{fig-1}. Notably, conservation of the total electronic charge does \textit{not}
imply conservation of the bichargon number. As bosonic quasiparticles, bichargons can be created or annihilated through charge-conserving pair processes involving oppositely
charged bichargons, or through charge exchange with other charged degrees
of freedom, such as conduction electrons or external reservoirs, analogous to the nonconservation
of magnon number in antiferromagnets. As we show below, the nonconserved quasiparticle nature of bichargons gives rise to transport properties qualitatively distinct from those of conventional charge carriers such as electrons.

\section{Response of bichargons to external perturbations}

We now explore the nonequilibrium transport and dynamical properties of bichargons under external perturbations, including temperature gradients and static and ac electric fields. We first show that, under a temperature gradient, thermally excited
incoherent bichargons in the two oppositely charged branches, when unequally
populated, generate a net charge current through their diffusion, giving
rise to a bichargon Seebeck effect. In contrast, despite carrying finite electric charges, bichargons, as nonconserved quasiparticles, do not support conventional dc electrical conduction under a static electric field: in the steady state, the field-driven drift current is compensated by the diffusion current arising from the field-induced spatial inhomogeneity of the bichargon density. We then show that an ac
electric field can parametrically generate coherent pairs of oppositely
charged bichargons. In the presence of a dc bias, these parametrically
excited bichargons support a finite dc charge current, providing a
bosonic analogue of bias-assisted photoconductivity mediated by photoexcited
electron--hole pairs in semiconductors.

\begin{figure}
    \centering
    \includegraphics[width=8.6 cm]{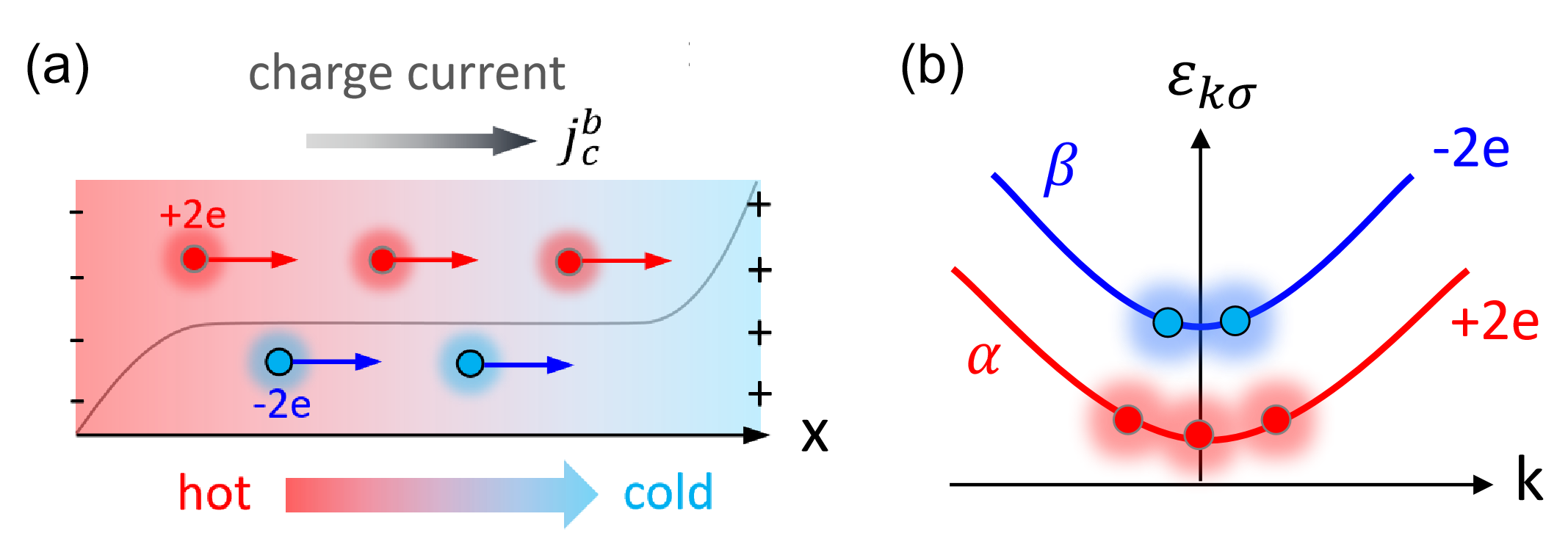}
    \caption{Bichargon Seebeck effect. (a) When thermally excited bichargons are
unequally populated in the two branches, their diffusion under a temperature
gradient generates a net charge current and opposite charge accumulation at
the two sample edges, constituting the charge counterpart of the magnonic
spin Seebeck effect in antiferromagnets. (b) Unequal thermal populations of the two nondegenerate bichargon branches
that arise when the chemical potential is tuned away from the
charge-neutrality point.}
    \label{fig-2}
\end{figure}

\subsection{Bichargon Seebeck effect}
We first consider the thermoelectric transport of bichargons in the presence of a temperature gradient ($\boldsymbol{\nabla}T$) and a static electric field ($\mathbf{E}$). A temperature gradient drives the diffusion of thermally excited bichargons and hence a charge current. At the particle-hole-symmetric point $h_z=0$, the two oppositely charged bichargon branches are degenerate and equally populated, such that their thermally driven charge currents exactly cancel. For $h_z\neq0$, however, the degeneracy is lifted and the two branches acquire unequal thermal populations, resulting in a finite net charge current $\mathbf{j}_c^{b}\propto\boldsymbol{\nabla}T$, as illustrated in Fig.~\ref{fig-2}. This bichargon-mediated thermoelectric response constitutes the charge-sector counterpart of the magnonic spin Seebeck effect in antiferromagnets~\cite{PhysRevB.87.014423,PhysRevLett.116.097204,PhysRevB.93.014425}. In a bulk charge-ordered system, the semiclassical Boltzmann equation for
the bichargon distribution function reads
\begin{equation}
\mathbf{v}_{\sigma}\cdot\frac{\partial f_{\sigma}}{\partial \mathbf{r}}+\frac{q_{\sigma}\mathbf{E}}{\hbar}\cdot\frac{\partial f_{\sigma}}{\partial\mathbf{k}}=-\frac{f_{\sigma}-f_{0}}{\tau_{\sigma}} \label{Bol}
\end{equation}
where $\mathbf{v}_{\sigma}
=\partial\varepsilon_{\mathbf{k}\sigma}/\partial(\hbar\mathbf{k})$
is the group velocity of bichargons in branch $\sigma$ with charge $q_{\sigma}$, while
$\hbar\dot{\mathbf{k}}=q_{\sigma}\mathbf{E}$
describes their semiclassical acceleration under the external electric field. The first and second terms on the left-hand side describe spatial diffusion and electric-field-driven drift in momentum space, respectively, while the collision term on the right-hand side relaxes the bichargon distribution toward the local equilibrium distribution $f_{0}$ with relaxation time $\tau_{\sigma}$. In contrast to electrons, bichargons are \textit{nonconserved} with a vanishing equilibrium chemical potential. Accordingly, their local equilibrium distribution obeys the Planck distribution,
\begin{equation}
f_{0}(\bar{\varepsilon}_{\mathbf{k}\sigma})=\left\{\exp\left(\frac{\bar{\varepsilon}_{\mathbf{k}\sigma}(\mathbf{r})}{k_{B}T(\mathbf{r})}\right)-1\right\}^{-1} \label{local}
\end{equation}
where 
$\bar{\varepsilon}_{\mathbf{k}\sigma}(\mathbf{r})
=\varepsilon_{\mathbf{k}\sigma}
-q_{\sigma}\mathbf{r}\cdot\mathbf{E}$
is the local bichargon energy including the electrostatic potential energy
in the applied electric field. Eq.~(\ref{local}) shows that an applied electric field renders the local equilibrium bichargon distribution spatially nonuniform, in contrast to the spatially uniform carrier distribution in the bulk of a conventional conductor under a uniform electric field. The electric field therefore has two competing effects on bichargon transport: it directly drives bichargon drift through the electric force $q_{\sigma}\mathbf{E}$, while simultaneously inducing a spatial variation of the local equilibrium distribution and hence a diffusive current, similar
to that driven by a temperature gradient. In the linear-response regime, solving Eq.~(\ref{Bol}) leads to
\begin{equation}
f_{\sigma}
\simeq
f_{0}(\bar{\varepsilon}_{\mathbf{k}\sigma})
-
\tau_{\sigma}
\mathbf{v}_{\sigma}\cdot\boldsymbol{\nabla}T
\frac{\partial f_{0}(\varepsilon_{\mathbf{k}\sigma})}{\partial T}.
\label{eq:flinear}
\end{equation}
Here, the electric-field-driven drift is exactly compensated by diffusion arising from the field-induced spatial inhomogeneity of the local equilibrium bichargon distribution, such that the electric field modifies only the local equilibrium distribution
and does not contribute to the nonequilibrium correction to $f_{\sigma}$. The resulting bulk bichargon charge current reads
\begin{align}
\mathbf{j}_{c}^{b}= \int\frac{d\mathbf{k}}{(2\pi)^{d}}\sum_{\sigma}q_{\sigma}\mathbf{v}_{\sigma} f_{\sigma}=-\mathcal{L}_{b} \nabla T 
\end{align}
where $\mathcal{L}_{b}$ is the bichargon thermoelectric (Seebeck) coefficient given by
\begin{equation}
\mathcal{L}_{b}=\int\frac{d\mathbf{k}}{(2\pi)^{d}}\sum_{\sigma}q_{\sigma}\tau_{\sigma} (\mathbf{v}_{\sigma}\cdot\mathbf{e}_{T})^2 \frac{\partial f_{0}(\varepsilon_{\mathbf{k}\sigma})}{\partial T}, \label{Seebeck}
\end{equation}
with $\mathbf{e}_{T}$ denoting the direction of the temperature gradient. As noted above, a static electric field cannot drive a steady bichargon
charge current because of the exact compensation between the drift and
diffusion currents. Therefore, despite the presence of thermally excited charged bichargons at finite temperatures, the charge-ordered state remains
electrically \textit{insulating} in the absence of extrinsic carrier doping. A temperature gradient, by contrast, may generate a finite bichargon charge
current. At $h_z=0$, the two oppositely charged bichargon branches are
degenerate and equally populated, such that their contributions to
Eq.~(\ref{Seebeck}) cancel. For $h_z\neq0$, however, the two branches are
split and acquire unequal equilibrium populations, resulting in a finite
thermoelectric coefficient. This bichargon thermoelectric response is
analogous to the magnonic spin Seebeck effect in an antiferromagnet under an
external magnetic field~\cite{PhysRevB.87.014423,PhysRevLett.116.097204,
PhysRevB.93.014425}.

The bulk bichargon charge current driven by a temperature gradient leads to
charge accumulation near the sample boundaries. In contrast to the charge
current carried by electrons, the bichargon charge current is not conserved
because bichargons, as quasiparticle excitations, can relax by transferring their charge to other charged degrees of freedom within the system or with
external reservoirs. For example, bichargons may
relax through charge transfer to mobile carriers in the surrounding
environment, accompanied by electrostatic screening.~Analogous to magnon spin diffusion in (anti)ferromagnets~\cite{
PhysRevLett.109.096603,PhysRevB.86.214424,PhysRevB.94.014412,PhysRevB.93.054412}, the resulting
bichargon charge accumulation obeys a continuity equation,
\begin{align}
\partial_{t}\rho_{c}^b+\boldsymbol{\nabla}\cdot \mathbf{j}_{c}^{b}=-\frac{\rho_{c}^{b}}{\tau_{b}},
\label{diff1}
\end{align}
where $\rho_c^b=\sum_{\sigma}q_{\sigma}\delta n_{\sigma}$ is the
nonequilibrium charge density, with $\delta n_{\sigma}$ the
bichargon number accumulation in branch $\sigma$. The bichargon charge
current is given by
\begin{equation}
\mathbf{j}_{c}^{b}
=
-\mathcal{L}_{b}\boldsymbol{\nabla}T
-\mathcal{D}_{b}\boldsymbol{\nabla}\rho_{c}^{b},
\label{diff2}
\end{equation}
where the second term describes the diffusive current driven by the
inhomogeneous bichargon charge accumulation, and
$\mathcal{D}_{b}$ is the bichargon diffusion coefficient. The relaxation term in Eq.~(\ref{diff1}) describes the decay of the bichargon charge density with a characteristic time $\tau_b$, arising, for
example, from charge transfer to other charged degrees of freedom within
the system or to external reservoirs, as discussed above. For a uniform
temperature gradient, Eq.~(\ref{diff1}) reduces in the steady state to
\begin{equation}
\nabla^2\rho_c^b
=
\frac{\rho_c^b}{\lambda_b^2},
\label{eq:charge_diffusion}
\end{equation}
where $\lambda_b=\sqrt{\mathcal{D}_b\tau_b}$ is the bichargon diffusion length. For a temperature gradient along the $x$ axis, as shown in Fig.~\ref{fig-2}, imposing the
vanishing-current condition at the boundaries $x=\pm L/2$ yields
\begin{align}
\rho_c^b(x)
=
-\frac{\mathcal{L}_b\lambda_b\nabla_{x}T}{\mathcal{D}_b}
\frac{
\sinh\left(x/\lambda_b\right)
}{
\cosh\left(L/2\lambda_b\right)
}\label{eq:charge_density}
\end{align}
where $L$ is the sample length along the temperature-gradient direction.
Opposite bichargon charges accumulate at the two boundaries over the
diffusion length $\lambda_b$, while the charge current is spatially symmetric
and for $L\gg\lambda_b$ approaches its bulk value $j_c^b=-\mathcal{L}_b\partial_xT$ in the sample interior.

Eq.~(\ref{eq:charge_density}) resemble the spin accumulation and current in magnets generated by the magnonic spin Seebeck effect~\cite{PhysRevB.89.014416,rezende2018magnon}. When the charge-ordered material is interfaced with a conductor, the nonequilibrium bichargon charge accumulated near the interface may be transferred to electronic carriers in the conductor, generating an electronic charge current or voltage, analogous to the conversion of magnon spin accumulation into an electronic
spin current across a magnet--metal interface~\cite{PhysRevLett.116.097204,
PhysRevB.89.014416,rezende2018magnon}. In practice, the bichargon-induced voltage may coexist with the conventional electronic Seebeck signal. The two contributions may, however, be distinguished by their different dependence on external control parameters. Remarkably, the bichargon Seebeck coefficient reverses sign upon reversing
$h_z=2\mu-U-2ZV$, which can be controlled by tuning the chemical potential
in charge-ordered materials with a gate voltage. This sign reversal provides
a characteristic signature for distinguishing the bichargon contribution
from the electronic background. Moreover, since bichargons exist only in the
charge-ordered state, the temperature dependence of the Seebeck voltage
across the charge-order transition provides an additional experimental probe
of bichargon-mediated charge transport.

\subsection{Electric-field effects on bichargons}
An electric field couples directly to bichargons through the electrostatic
potential energy $-q_{\sigma}\mathbf{r}\cdot\mathbf{E}$. Unlike conventional electrons, however, such a monopolar coupling cannot drive a steady bichargon charge current because of the nonconserved quasiparticle nature of bichargons, as demonstrated above. When the charge-ordered state simultaneously breaks inversion symmetry, the asymmetric charge distribution can give rise to a finite electric polarization, as realized, for example, in electronic ferroelectrics~\cite{PhysRevB.54.17452,ikeda2005ferroelectricity,ishihara2010electronic}. Beyond the monopolar coupling, an electric field $E$ applied along the polarization direction can therefore couple to the charge order and hence to the staggered pseudospins through the dipolar Stark coupling~\cite{tang2026pseudospin},
\begin{align}
H_{\text{ext}}
=
-p_0E
\left(
\sum_{i\in A}\tau_i^z
-
\sum_{j\in B}\tau_j^z
\right),
\label{eq:HE}
\end{align}
where $p_0$ denotes the intracell dipole moment associated with the charge order and characterizes the coupling strength. Below, we investigate the
effects of the dipolar coupling in Eq.~(\ref{eq:HE}) on bichargons under
static and alternating (ac) electric fields.

\begin{figure}
    \centering
    \includegraphics[width=8.2 cm]{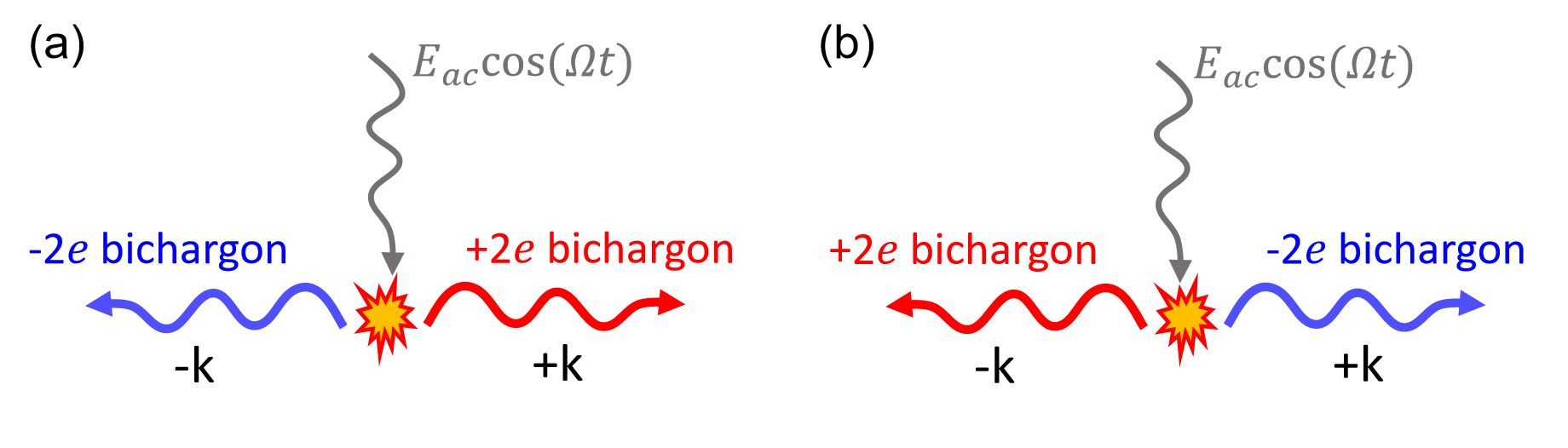}
    \caption{Parametric excitation of coherent pairs of oppositely charged bichargons
propagating in opposite directions via (a) the process
$\alpha_{\mathbf{k}}^{\dagger}\beta_{-\mathbf{k}}^{\dagger}$ in
Eq.~(\ref{eq:HRWA}) and (b) its inversion-related counterpart. The two equivalent processes carry equal and opposite charge currents,
resulting in a vanishing macroscopic current, whereas an applied dc bias
lifts this cancellation and drives a net dc charge current carried by the
parametrically generated bichargon pairs, analogous to photoconductivity in semiconductors mediated by photoexcited electron--hole pairs.}
    \label{fig-parametric}
\end{figure}

 \subsubsection{Static electric field}
We first consider the dipolar coupling to a static electric field $E_{\rm st}$. Applying the Holstein--Primakoff transformation to Eq.~(\ref{eq:HE}) yields
\begin{equation}
H_{\text{ext}}
=p_{0}E_{\text{st}}
\sum_{\mathbf{k}}
\left(
a_{\mathbf{k}}^\dagger a_{\mathbf{k}}
+
b_{\mathbf{k}}^\dagger b_{\mathbf{k}}
\right) +\text{const.}
\label{eq:HEboson}
\end{equation}
where the constant term represents the field-induced shift of the
ground-state energy. Including Eq.~(\ref{eq:HEboson}) amounts to the replacement
$\mathcal{A}\rightarrow\mathcal{A}+p_0E_{\rm st}$ in
Eq.~(\ref{eq:Hpm_quad}), yielding the bichargon dispersions
\begin{equation}
\varepsilon_{\mathbf{k}\alpha(\beta)}
=
\sqrt{
\left(Z\tau_0J_z+p_{0}E_{\text{st}}\right)^2
-
\left(
Z\tau_0J_\perp\gamma_{\mathbf{k}}
\right)^2
}\pm h_{z},
\label{eq:Edisp}
\end{equation}
where the static electric field acts as an effective renormalization of the
Ising-type pseudospin coupling $J_z$. For $E_{\rm st}>0$, the field is aligned with the charge-order-induced
intracell dipole and stabilizes the charge-ordered state by increasing
the bichargon gap. Conversely, for $E_{\rm st}<0$, the field opposes the
intracell dipole and reduces the bichargon gap, thereby destabilizing the
charge order.

The electric-field dependence of the bichargon dispersion implies that
bichargons also carry an intrinsic electric dipole moment when the charge
order simultaneously breaks inversion symmetry, similar to ferrons in
ferroelectrics~\cite{Tang2022,Bauer2022,PhysRevApplied.20.050501,Tang2024}.
The corresponding dipole moment can be identified from the field
derivative of the bichargon energy as~\cite{Tang2022,gm7n-8ftp}
\begin{equation}
d_{\mathbf{k}\sigma}
=
-\left.
\frac{\partial\varepsilon_{\mathbf{k}\sigma}}
{\partial E_{\rm st}}
\right|_{E_{\rm st}=0}
=
-\frac{p_0J_z}
{\sqrt{J_z^2-J_\perp^2\gamma_{\mathbf{k}}^2}},
\label{eq:bichargon_dipole}
\end{equation}
which is identical for the two bichargon branches. The negative sign
indicates that the bichargon dipole moment is opposite to that of the
ground-state intracell dipole, meaning that bichargon excitations reduce
the electric polarization, as do ferrons in ferroelectrics. Unlike ferrons,
however, bichargons additionally carry electric charges of $\pm2e$.

The stability of the charge-ordered state requires the lower bichargon
branch to remain positive. A sufficiently strong electric field opposing
the charge-order polarization can destabilize the initial charge-ordered
state and induce reversal of the charge-order polarity and the associated
electric polarization. The instability sets in when the uniform
($\mathbf{k}=0$) lower bichargon branch in Eq.~(\ref{eq:Edisp}) softens to zero, yielding the
critical field
\begin{equation}
E_{\text{st}}^{c}
=
\frac{-Z\tau_0J_z
+
\sqrt{
(Z\tau_0J_\perp)^2+h_z^2
}}{p_{0}},
\label{eq:combinedhc}
\end{equation}
which in the absence of $h_z$ reduces to the result reported in Ref.~\cite{tang2026pseudospin}. For fields beyond the instability threshold, the lower branch in Eq.~(\ref{eq:Edisp}) becomes negative, signaling an instability of the initial charge-ordered state. The system is then driven toward the reversed charge-ordered state with the staggered pseudospin configuration $\langle\tau_A^z\rangle=-\tau_0$ and $\langle\tau_B^z\rangle=\tau_0$. Performing the Holstein--Primakoff expansion about this reversed
configuration corresponds to $\tau_0\rightarrow-\tau_0$ in Eq.~(\ref{eq:Edisp}) and yields positive
excitation energies for both bichargon branches, indicating that the reversed
charge-ordered state is stable against pseudospin fluctuations.

\subsubsection{Parametric excitation by an ac electric field}
We next consider the parametric excitation of bichargons by a time-dependent electric field, $E(t)=E_{\rm ac}\cos\Omega t$, with amplitude $E_{\rm ac}$ and frequency $\Omega$. When the driving frequency approaches the sum of the frequencies of the two bichargon branches, $\Omega\simeq\omega_{\mathbf{k}\alpha}+\omega_{-\mathbf{k}\beta}$, the ac
field resonantly creates and annihilates bichargon pairs. Retaining only the
resonant terms within the rotating-wave approximation, the dipolar coupling
in Eq.~(\ref{eq:HE}) becomes
\begin{equation}
H_{\rm par}
=
-\hbar
\sum_{\mathbf{k}}
\left[g_{\mathbf{k}}
\alpha_{\mathbf{k}}^\dagger
\beta_{-\mathbf{k}}^\dagger
e^{-i\delta_{\mathbf{k}}t}
+H.c.
\right],
\label{eq:HRWA}
\end{equation}
where $\delta_{\mathbf{k}}
=
\Omega
-
\omega_{\mathbf{k}\alpha}
-
\omega_{-\mathbf{k}\beta}$ is the detuning, and $g_{\mathbf{k}}
=p_0E_{\rm ac}u_{\mathbf{k}}v_{\mathbf{k}}/\hbar$
is the parametric coupling strength. Here, we have used the interaction-picture time evolution for the bichargon operators, $\alpha_{\mathbf{k}}(t)
=
\alpha_{\mathbf{k}}
e^{-i\varepsilon_{\mathbf{k}\alpha}t/\hbar}$ and $\beta_{\mathbf{k}}(t)
=
\beta_{\mathbf{k}}
e^{-i\varepsilon_{-\mathbf{k}\beta}t/\hbar}$.

Eq.~(\ref{eq:HRWA}) takes the form of a two-mode squeezing Hamiltonian~\cite{WallsMilburn2008} and describes the parametric excitation of a coherent pair of $\alpha_{\mathbf{k}}$ and $\beta_{-\mathbf{k}}$ bichargons, as shown in Fig.~\ref{fig-parametric}. Since the two bichargons carry opposite charges, the pair excitation conserves both the total charge and crystal momentum of the system. Starting from the bichargon vacuum, the populations of the two modes evolve as
\begin{equation}
n_{\alpha,\mathbf{k}}(t)
=
n_{\beta,-\mathbf{k}}(t)
=
\frac{g_{\mathbf{k}}^2}{\Lambda_{\mathbf{k}}^2}
\sinh^2\left(\Lambda_{\mathbf{k}}t\right),
\label{eq:parametric_population}
\end{equation}
with $\Lambda_{\mathbf{k}}
=
\sqrt{g_{\mathbf{k}}^2-\delta_{\mathbf{k}}^2/4}$. For $|\delta_{\mathbf{k}}|<2|g_{\mathbf{k}}|$,
$\Lambda_{\mathbf{k}}$ is real, leading to exponential parametric
amplification of the bichargon population. At exact resonance ($\delta_{\mathbf{k}}=0$), Eq.~(\ref{eq:parametric_population}) reduces to
$n_{\alpha,\mathbf{k}}(t)=n_{\beta,-\mathbf{k}}(t)=\sinh^2(\vert g_{\mathbf{k}}\vert t)$. The equal populations of the two branches reflect their pairwise generation, whereby each $+2e$ bichargon is created together with a $-2e$ bichargon of opposite momentum.

To account phenomenologically for relaxation, we introduce an equal damping rate $\Gamma_{\mathbf{k}}$ for the two bichargon modes. The equations of motion generated by Eq.~(\ref{eq:HRWA}) then become
\begin{equation}
\frac{d}{dt}
\begin{pmatrix}
\tilde{\alpha}_{\mathbf{k}}\\
\tilde{\beta}_{\mathbf{k}}
\end{pmatrix}
=
\begin{pmatrix}
-\Gamma_{\mathbf{k}}+i\delta_{\mathbf{k}}/2
&
ig_{\mathbf{k}}
\\
-ig_{\mathbf{k}}
&
-\Gamma_{\mathbf{k}}-i\delta_{\mathbf{k}}/2
\end{pmatrix}
\begin{pmatrix}
\tilde{\alpha}_{\mathbf{k}}\\
\tilde{\beta}_{\mathbf{k}}
\end{pmatrix}\label{para}
\end{equation}
where $
\tilde{\alpha}_{\mathbf{k}}
=
\alpha_{\mathbf{k}}
e^{i\delta_{\mathbf{k}}t/2} $ and $
\tilde{\beta}_{\mathbf{k}}
=
\beta_{-\mathbf{k}}^\dagger
e^{-i\delta_{\mathbf{k}}t/2}$
are slowly varying operators in the rotating frame. The corresponding
growth exponents of the coupled mode amplitudes are
$\lambda_{\pm}=-\Gamma_{\mathbf{k}}\pm\Lambda_{\mathbf{k}}$.
Parametric amplification occurs when the parametric gain exceeds the damping, i.e.,
$\Gamma_{\mathbf{k}}<\Lambda_{\mathbf{k}}$. At exact resonance, this condition reduces to $\vert g_{\mathbf{k}}\vert >\Gamma_{\mathbf{k}}$, yielding the threshold ac electric-field amplitude
\begin{equation}
E_{\rm ac}^{c}
=
\frac{
2\hbar\varepsilon_{\mathbf{k}}\Gamma_{\mathbf{k}}}{
p_0 Z\tau_0J_\perp|\gamma_{\mathbf{k}}|}
\label{eq:threshold}
\end{equation}
where we have used $|u_{\mathbf{k}}v_{\mathbf{k}}|
=
Z\tau_0J_\perp|\gamma_{\mathbf{k}}|/(2\varepsilon_{\mathbf{k}})$. Eq.~(\ref{eq:threshold}) implies that, for a given driving frequency, the bichargon pair with the wave vector that minimizes $\varepsilon_{\mathbf{k}}\Gamma_{\mathbf{k}}/|\gamma_{\mathbf{k}}|$ on the resonance shell $\varepsilon_{\mathbf{k}\alpha}+\varepsilon_{-\mathbf{k}\beta}=\hbar\Omega$ is the first to become parametrically excited. Frequency-resolved
parametric excitation therefore provides a spectroscopic probe of the
bichargon dispersion, analogous to two-magnon parametric excitation in
magnets~\cite{suhl1957theory,bracher2017parallel}.

While parametric pumping conserves both the total electric charge and crystal momentum, an individual bichargon-pair excitation,
\begin{equation}
\alpha_{\mathbf{k}}^{\dagger}\beta_{-\mathbf{k}}^{\dagger}:
\quad
|\mathrm{vac}\rangle
\rightarrow
(+2e,\mathbf{k})_{\alpha}
+
(-2e,-\mathbf{k})_{\beta},
\end{equation}
causes a charge current, since the two oppositely charged bichargons propagate in opposite directions. The inversion-related process $\alpha_{-\mathbf{k}}^{\dagger}\beta_{\mathbf{k}}^{\dagger}:|\mathrm{vac}\rangle
\rightarrow
(+2e,-\mathbf{k})_{\alpha}
+
(-2e,\mathbf{k})_{\beta}$, however, carries an equal and opposite current, as illustrated in
Fig.~\ref{fig-parametric}, where $|\mathrm{vac}\rangle$ denotes the bichargon vacuum. In an inversion-symmetric system, a spatially uniform ac field excites the two processes with equal weight, resulting in a vanishing macroscopic charge current.

A dc bias electric field lifts the cancellation between the two inversion-related processes by driving oppositely charged bichargons in opposite directions, thereby converting the pumped bichargon population into a finite dc charge current. This response is qualitatively distinct from that of thermally excited bichargons under a static electric field. In the latter case, the field-induced drift current is accompanied by an opposing diffusion current arising from the field-induced spatial redistribution of the bichargon density, such that a thermal bichargon gas remains electrically insulating in the steady state. Parametric pumping, by contrast, continuously injects a spatially uniform nonequilibrium population of correlated bichargon pairs that support a finite drift current under an applied dc bias. This mechanism is analogous to photoconductivity in semiconductors mediated by photoexcited electron--hole pairs~\cite{sze2021physics}, with the charge carriers here being parametrically excited bosonic bichargon pairs.

\section{Discussion and Conclusion}
Charged bosonic degrees of freedom have previously appeared in several
contexts in Hubbard-type models. In particular, the pseudospin symmetry of
the Hubbard model underlies $\eta$ pairing and associated collective
modes~\cite{Zhang1990,Betsuyaku1991}, while emergent charge-$2e$ bosonic
degrees of freedom have been discussed in low-energy descriptions of doped
Mott insulators~\cite{Leigh2007,Choy2008}. Although related through their
common origin in strong electronic correlations, these objects are
conceptually distinct from the bichargons considered here. Bichargons are
quantized collective excitations above an established charge-ordered
state, arising from transverse quantum fluctuations of the ordered charge
pseudospins, and occur as two branches carrying opposite electric charges
$+2e$ and $-2e$. They therefore do not represent preformed Cooper pairs or
a superconducting pairing field, but rather charge-carrying collective modes
of the charge order itself.

Bichargons should be distinguished from other charged excitations and
composite objects in quantum materials. Although their charge magnitude
coincides with that of Cooper pairs, their existence requires neither
superconducting phase coherence nor dissipationless charge transport. They
are also conceptually distinct from chargons and holons, which emerge from
electron fractionalization and carry charge of magnitude $e$ independently
of spin~\cite{SenthilFisher2000,LeeNagaosaWen2006}, as well as from
conventional amplitudon and phason modes of charge-density
waves~\cite{Gruner1988}, which classify collective fluctuations by the
amplitude and phase of the order parameter. In contrast, bichargons originate from local electron-pair transitions
between empty and doubly occupied states and resolve quantized
collective excitations of charge order according to the electric charge they
carry, providing a charge-sector counterpart of magnons in spin-ordered
states. Despite carrying quantized electric charge, their quasiparticle
number is not conserved. This distinction from conventional electronic carriers leads to an
unusual dichotomy in their response to external perturbations: a thermal
bichargon gas can transport charge under a temperature gradient, giving rise
to a bichargon Seebeck effect, while remaining electrically insulating in
response to an applied static electric field. Bichargons can therefore
mediate nonequilibrium charge transport without rendering the underlying
charge-ordered state electrically conducting.

The charged nature of bichargons further enables electrical control of their
spectrum and population. When the charge order also breaks inversion
symmetry, bichargons possess an intrinsic electric dipole moment in addition
to their quantized monopolar charge, enabling electric-field tunability of
their spectrum, similar to ferrons in ferroelectrics~\cite{Tang2022,Bauer2022,PhysRevApplied.20.050501,Tang2024}. An ac electric field can parametrically generate correlated bichargon pairs
with opposite charges and momenta through dynamical Stark coupling near the
two-bichargon resonance, analogous to parallel parametric magnon-pair
generation in magnetic systems~\cite{suhl1957theory,bracher2017parallel}. Although each generated pair is charge neutral, its oppositely charged
constituents contribute additively to the charge current when driven in
opposite directions. A dc bias can therefore convert the pumped bichargon
population into a finite electrical current, providing a bosonic counterpart
of semiconductor photoconductivity mediated by photoexcited electron--hole
pairs. Parametric excitation thus provides a route to generating mobile
charged collective excitations in an otherwise insulating charge-ordered
state.

In conclusion, we have predicted two branches of bichargons in
charge-ordered systems that carry opposite quantized electric charges
$\pm 2e$, closely paralleling the two spin-carrying magnon branches of
a bipartite antiferromagnet. As bosonic quasiparticles with a nonconserved
particle number, bichargons exhibit charge transport and nonequilibrium
responses qualitatively distinct from those of conventional electronic
carriers, including thermally driven charge transport without conventional
dc electrical conduction. Beyond thermoelectric transport, bichargons can
be parametrically generated by an ac electric field, while their spectrum
can be tuned by a gate electric field, providing complementary experimental
avenues for probing their existence and characteristic properties. These
results establish bichargons as a distinct class of charged collective
excitations and open a route toward exploiting collective charge degrees
of freedom for electrical transport, spectroscopy, and nonequilibrium
control in charge-ordered quantum materials.

\begin{acknowledgments}
This work was supported by JSPS KAKENHI Grant-in-Aid for Scientific Research (B)
(Grant No. 26K00625).
\end{acknowledgments}

\bibliography{reference}

\end{document}